\documentclass[aps,english,showpacs,twocolumn]{revtex4-2}
\usepackage{amsfonts}
\usepackage{amssymb}
\usepackage{amsmath}
\usepackage{graphicx}
\usepackage{epsfig}
\usepackage{subfigure}
\usepackage{color}
\usepackage{amsmath,bm}
\usepackage{booktabs}
\usepackage{appendix}
\usepackage[colorlinks=true, linkcolor=blue, citecolor=blue, urlcolor=blue]{hyperref}

\begin{document}

\title{Quantum Many-Body Scars, Magnon-Pair Condensation, and Hilbert Space
Fragmentation in an Anisotropic Heisenberg Model}
\author{J. Y. Liu-Sun}
\author{Z. Song}
\email{songtc@nankai.edu.cn}

\begin{abstract}
We investigate a spin-$1/2$ anisotropic Heisenberg model on a lattice
consisting of two identical bipartite sublattices. A family of exact
eigenstates generated by the restricted spectrum generating algebra (RSGA)
constitutes quantum many-body scar states, characterized by subextensive
entanglement entropy and supporting. These scar states are magnon-pair
condensates exhibiting off-diagonal long-range order (ODLRO). At the
resonance point of the inter-sublattice interaction, the model exactly maps
onto a mixed spin-$1$ and spin-$0$ XY model on a bipartite lattice, which
decomposes into independent sub-Hamiltonians labeled by all possible spin
configurations. Each spin-$0$ particle is dynamically isolated from its
neighbors and acts as a kinetic constraint, giving rise to emergent Hilbert
space fragmentation (HSF). Our work establishes an exactly solvable platform
in which quantum many-body scars, magnon-pair condensation exhibiting
off-diagonal long-range order, and Hilbert space fragmentation naturally
coexist.
\end{abstract}

\affiliation{School of Physics, Nankai University, Tianjin 300071, China}
\maketitle

\section{Introduction}

\label{Introduction}

Understanding the mechanisms responsible for ergodicity breaking in isolated
quantum many-body systems has become a central topic in nonequilibrium
physics. While generic interacting systems are expected to thermalize
according to the eigenstate thermalization hypothesis (ETH) \cite%
{Deutsch1991,Srednicki1994,Rigol2008}, a growing number of exceptions have
been discovered. Prominent examples include integrable systems \cite%
{Vidmar2016}, many-body localized phases \cite{Nandkishore2015,Abanin2019},
quantum many-body scars (QMBS) \cite%
{Turner2018,Serbyn2021,Moudgalya2022Fragmentation}, and Hilbert-space
fragmentation (HSF) \cite{Khemani2020,Moudgalya2022Fragmentation}. These
phenomena reveal that nonthermal behavior can emerge even in the absence of
conventional conservation laws and provide new routes toward stabilizing
coherent quantum dynamics.

Quantum many-body scars are atypical eigenstates embedded in an otherwise
thermal spectrum \cite{Turner2018}. Such states possess anomalously low
entanglement entropy and give rise to long-lived oscillatory dynamics from
special initial states \cite{Turner2018,Serbyn2021}. Since their discovery
in kinetically constrained Rydberg atom systems \cite{Bernien2017,Turner2018}%
, scarred eigenstates have been identified in a variety of models, including
spin chains \cite{Shiraishi2017,Lin2019}, Hubbard models \cite%
{Moudgalya2020Hubbard}, and higher-spin systems \cite{Mark2020}. A
particularly fruitful approach to constructing exact scar states is based on
restric spectrum-generating algebras (SGAs), where towers of equally spaced
eigenstates arise from special ladder operators acting on simple reference
states \cite{Moudgalya2020Hubbard,Choi2019}.

Another mechanism for weak ergodicity breaking is Hilbert-space
fragmentation. In fragmented systems, the Hilbert space splits into
exponentially many dynamically disconnected sectors due to local constraints
or emergent conservation laws \cite{Sala2020,Khemani2020}. Unlike
conventional symmetry decomposition, fragmentation can generate an extensive
number of small invariant subspaces and produce atypical eigenstates with
low entanglement \cite{Moudgalya2022Fragmentation}. HSF has been observed in
a broad range of systems, including constrained spin models \cite%
{Sala2020,Khemani2020,He2026,Ma2026arXiv}, lattice gauge theories \cite%
{Lan2022}, and dipole-conserving quantum dynamics \cite%
{Pai2019,GuardadoSanchez2020}.

Although QMBS and HSF are often studied as distinct manifestations of weak
ergodicity breaking, recent developments suggest that they may share a
common origin in certain classes of constrained quantum systems \cite%
{Moudgalya2022Fragmentation}. Identifying simple microscopic models in which
both phenomena emerge naturally remains an important open problem.

In this work, we investigate a spin-$1/2$ anisotropic Heisenberg model
defined on a lattice consisting of two identical bipartite sublattices. Our
aim is to explore the interplay between quantum many-body scars,
off-diagonal long-range order, and Hilbert space fragmentation within a
unified and exactly solvable framework.

We first show that the restricted spectrum generating algebra (RSGA)
generates a family of exact eigenstates embedded in the many-body spectrum.
These states constitute quantum many-body scar states, characterized by
subextensive entanglement entropy and supporting periodic quantum revivals.
Furthermore, we demonstrate that they are magnon-pair condensates exhibiting
off-diagonal long-range order \cite{Yang1962}, thereby establishing a direct
connection between quantum many-body scarring and many-body condensation.

We then focus on the resonance point of the inter-sublattice interaction,
where the model admits an exact mapping onto a mixed spin-$1$ and spin-$0$
XY model on a bipartite lattice. The mapped Hamiltonian decomposes into
independent sub-Hamiltonians labeled by all possible spin configurations.
Since each spin-$0$ particle is dynamically isolated from its neighboring
spins, it acts as a kinetic constraint that naturally gives rise to Hilbert
space fragmentation.

Taken together, these results establish an exactly solvable quantum spin
model in which quantum many-body scars, magnon-pair condensation with
off-diagonal long-range order, and Hilbert space fragmentation naturally
coexist. Our work provides a unified setting for investigating the interplay
among these seemingly distinct nonequilibrium phenomena and offers new
insight into constrained quantum many-body dynamics.

The rest of this paper is organized as follows. In Section \ref{Model and
SGA}, we introduce the model Hamiltonian and present the $\eta$-pairing
symmetry of the system for a special case using the Jordan-Wigner
transformation. In Section \ref{Exact eigenstates as QMBS}, we construct a
family of magnon-pair eigenstates via the RSGA for the general case. In
Section \ref{Entanglement entropy and magnon-pair correlations}, we
investigate the scaling of the entanglement entropy and the magnon-pair
correlations. In Section \ref{Hilbert space fragmentation}, we investigate
the emergence of Hilbert space fragmentation at the resonance point.
Finally, we summarize our results and discuss their implications in Section %
\ref{summary}.

\section{Model and SGA}

\label{Model and SGA}

We consider the anisotropic Heisenberg model on a lattice composed of two
identical bipartite sublattices indexed by $m=1$ and $2$. The Hamiltonian
has the form%
\begin{equation}
H=\sum_{j\in A\cup B}V_{j}+\sum_{i\in A,j\in B}T_{ij},  \label{H}
\end{equation}%
where term $V_{j}$ describes the XXZ type coupling for a dimer 
\begin{equation}
V_{j}=J_{\perp }(s_{j,1}^{+}s_{j,2}^{-}+\mathrm{H.}\text{\textrm{c}}\mathrm{.})%
+J_{z}\left( s_{j,1}^{z}s_{j,2}^{z}-\frac{1}{4}\right) ,  \label{V_j}
\end{equation}%
and term $T_{ij}$ describes the XY type coupling between two dimers 
\begin{equation}
T_{ij}=J\sum_{m=1,2}s_{i,m}^{+}s_{j,m}^{-}+J_{\times
}(s_{i,1}^{+}s_{j,2}^{-}+s_{i,2}^{+}s_{j,1}^{-})+\mathrm{H.}\text{\textrm{c}}%
\mathrm{.}.  \label{T_ij}
\end{equation}%
\begin{figure}[t]
\centering
\includegraphics[width=0.48\textwidth]{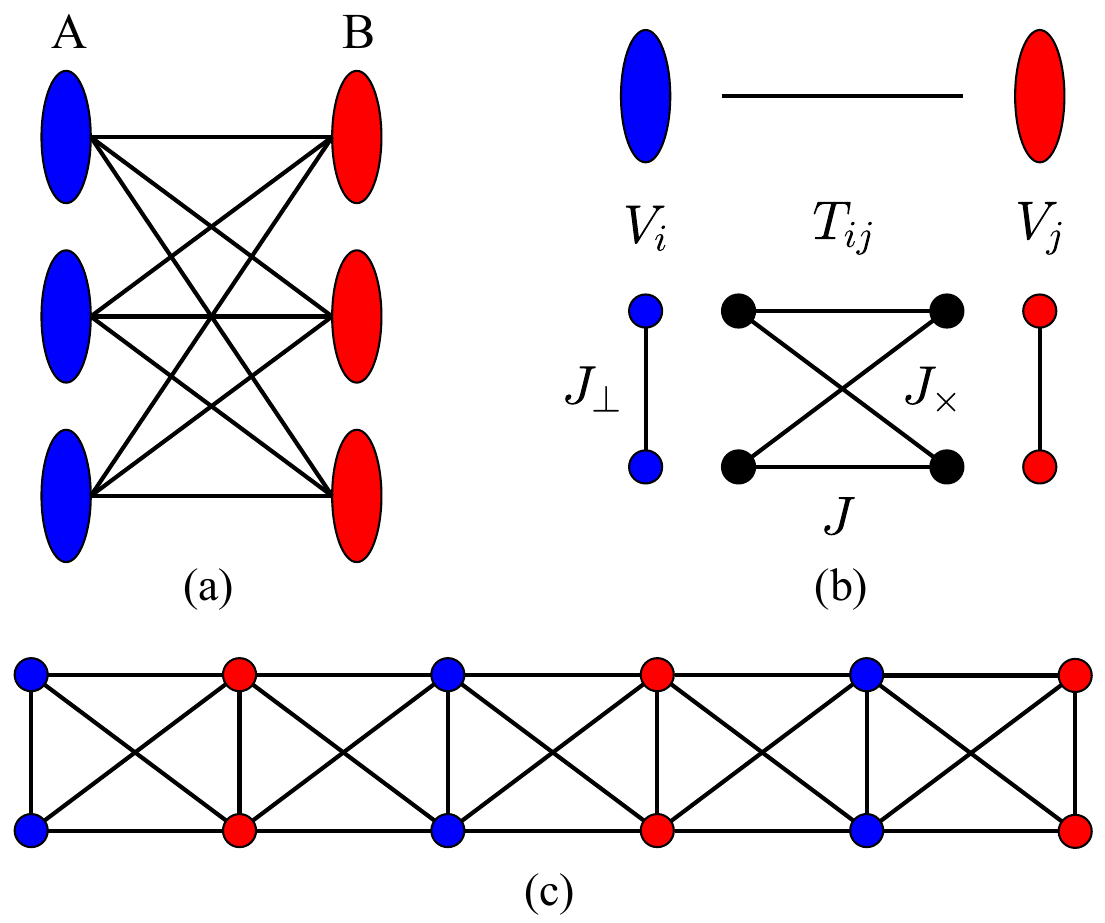}
\caption{Schematic illustration of the anisotropic Heisenberg model on a
lattice composed of two identical sublattices. (a) Generic bipartite lattice
geometry. The intra-dimer coupling $V_{j}$ acts on the two spins within each
dimer, while the inter-dimer coupling $T_{ij}$ connects dimers belonging to
different sublattices A and B. (b) Local dimer structure with anisotropic
couplings $J_{\perp }$ and $J_{z}$, given in Eqs. (\protect\ref{V_j}). (c)
Spin-ladder as an example of the model, given in Eqs. (\protect\ref{H_Ladd}
), where the inter-rung couplings are XY-type, denoted by $J$ and $J_{\times
}$, respectively.}
\label{fig1}
\end{figure}
Here ($s_{i,m}^{\pm },s_{i,m}^{z}$) are spin-$1/2$ operators at site $i$ in
the $m$-th ($m=1,2$) sublattice.\ Each dimer $V_{j}$\ acts as the
fundamental building block,\ while $T_{ij}$\ is the connection between pair
of dimers with $i\in A$ and $j\in B$.\ The lattice geometry is shown in Fig.~%
\ref{fig1}. The intra-sublattice and inter-sublattice interacting strengths
are $J$ and ($J_{\times },J_{\perp },J_{z}$), respectively. We introduce the
total spin operators 
\begin{eqnarray}
s^{+} &=&\left( s^{-}\right) ^{\dagger }=\sum_{j,m}s_{j,m}^{+}, \\
s^{z} &=&\sum_{j,m}s_{j,m}^{z},
\end{eqnarray}
which obey the SU(2) Lie algebra, i.e., $[s^{+},s^{-}]=2s^{z}$, and $%
[s^{z},s^{\pm }]=\pm s^{\pm }$. The Hamiltonian $s^{z}$\ conservative, i.e., 
$[s^{z},H]=0$.

A simplest example is a spin ladder system with open boundary condition,
with the Hamiltonian%
\begin{equation}
H_{\text{\textrm{Ladd}}}=\sum_{j=1}^{N}V_{j}+\sum_{j=1}^{N-1}T_{j(j+1)},
\label{H_Ladd}
\end{equation}%
which is schematically illustrated in Fig. \ref{fig1}. Quantum spin
ladders constitute one of the most important quasi-one-dimensional quantum
many-body systems. They have been experimentally realized in a variety of
condensed-matter compounds, including cuprate ladder materials, where
neutron scattering, nuclear magnetic resonance, and thermodynamic
measurements have revealed spin gaps, short-range singlet correlations, and
field-induced quantum phase transitions \cite{Dagotto1996,Giamarchi2008,Lake2010}. More recently, rapid progress in
quantum simulation has enabled highly controllable realizations of ladder
Hamiltonians \cite{Atala2014,Hirthe2023}. 

In the following, we will investigate this concrete system in order to
demonstrate the main idea of this work. We consider the case with $J_{\times
}=J_{\perp }=0$ to simplify the analysis. We apply the Jordan-Wigner
transformation on each sublattice 
\begin{eqnarray}
s_{j,m}^{+} &=&(s_{j,m}^{-})^{\dagger }=c_{j,m}^{\dagger }\exp (i\pi
\sum_{l<j}c_{l,m}^{\dagger }c_{l,m}),  \notag \\
s_{j,m}^{z} &=&\frac{1}{2}[s_{j,m}^{+},s_{j,m}^{-}]=c_{j,m}^{\dagger
}c_{j,m}-\frac{1}{2},
\end{eqnarray}%
to express the spin operators in terms of fermion creation and annihilation
operators $c_{j,m}^{\dagger }$ and $c_{j,m}$, satisfying $%
\{c_{j,m},c_{l,n}^{\dagger }\}=\delta _{j,l}\delta _{m,n}$ and $%
\{c_{j,m},c_{l,n}\}=0$. Thus the Hamiltonian $H$ can be mapped into the
explicit form%
\begin{eqnarray}
H_{\text{\textrm{Ladd}}} &=&J\sum_{j,m}c_{j,m}^{\dagger }c_{j+1,m}+\mathrm{H.}%
\text{\textrm{c}}  \notag \\
&&+J_{z}\sum_{j}(c_{j,1}^{\dagger }c_{j,1}-\frac{1}{2})(c_{j,2}^{\dagger
}c_{j,2}-\frac{1}{2}),
\end{eqnarray}%
where a constant is neglected. Obviously, $H_{\text{\textrm{Ladd}}}$\ is an
interacting two-component fermionic system on a bipartite lattice.

We introduce a set of pseudo-spin operators

\begin{eqnarray}
\eta ^{+} &=&\left( \eta ^{-}\right) ^{\dag }=\sum_{j}\left( -1\right)
^{j}c_{j,1}^{\dagger }c_{j,2}^{\dagger },  \notag \\
\eta ^{z} &=&\frac{1}{2}\sum_{j}\left( c_{j,1}^{\dagger
}c_{j,1}+c_{j,2}^{\dagger }c_{j,2}-1\right) ,
\end{eqnarray}%
which obey the SU(2) Lie algebra, i.e., $[\eta ^{+},\eta ^{-}]=2\eta ^{z}$,
and $[\eta ^{z},\eta ^{\pm }]=\pm \eta ^{\pm }$. Indeed, $H_{\text{\textrm{%
Ladd}}}$ is a Fermi-Hubbard model, provided we identify the $m=1$ sublattice
as spin-up and the $m=2$ sublattice as spin-down. At this point, $\eta ^{\pm
}$ corresponds to the $\eta $-pairing operators, and the Hamiltonian
respects the inherent SU(2) symmetry 
\begin{equation}
\left[ H_{\text{\textrm{Ladd}}},\eta ^{\pm }\right] =\left[ H_{\text{\textrm{%
Ladd}}},\eta ^{z}\right] =0.  \label{eta-symmetry}
\end{equation}%
\ This allows us to construct a set of eigenstates in the form $\left( \eta
^{+}\right) ^{n}\left\vert \text{\textrm{Vac}}\right\rangle $ based on the
spectrum generating algebra (SGA) \cite{Yang1989}, where $\left\vert \text{%
\textrm{Vac}}\right\rangle $ is the vacuum state of the fermions.

This result provides us three implications. (i) The original spin ladder
system has a hidden symmetry associated with a complicated operator%
\begin{eqnarray}
\eta ^{+} &=&\left( \eta ^{-}\right) ^{\dag }  \notag \\
&=&\sum_{j}\left( -1\right) ^{j}\exp [i\pi \sum_{l<j}\left(
s_{l,1}^{z}+s_{l,2}^{z}+1\right) ]s_{j,1}^{+}s_{j,2}^{+}  \notag \\
&=&-\sum_{j}\exp [i\pi \sum_{l<j}\left( s_{l,1}^{z}+s_{l,2}^{z}\right)
]s_{j,1}^{+}s_{j,2}^{+},
\end{eqnarray}%
which obtained by the inversed Jordan-Wigner transformation. (ii) We note
that 
\begin{eqnarray}
(s_{j,1}^{+}s_{j+1,2}^{-}+s_{j+1,1}^{+}s_{j,2}^{-})\left( \eta ^{+}\right)
^{n}\left\vert \text{\textrm{Vac}}\right\rangle &=&0,  \notag \\
s_{j,1}^{+}s_{j,2}^{-}\left( \eta ^{+}\right) ^{n}\left\vert \text{\textrm{%
Vac}}\right\rangle &=&0,
\end{eqnarray}%
which indicates that the state $\left( \eta ^{+}\right) ^{n}\left\vert \text{%
\textrm{Vac}}\right\rangle $\ is still the eigenstate of the Hamiltonian $H_{%
\text{\textrm{Ladd}}}$\ with nonzero $J_{\times }$ and $J_{\perp }$,
although the $\eta $-symmetry is broken in this situation. It motives us to
seek the existence of magnon-pair eigenstates for the original Hamiltonian.

\section{Exact eigenstates as QMBS}

\label{Exact eigenstates as QMBS}

In this section, we focus on the pairing states of the original spin
Hamiltonian (Eq. (\ref{H})). We introduce a set of magnon-pair operators

\begin{eqnarray}
Q^{+} &=&\left( Q^{-}\right) ^{\dag }=\sum_{i\in
A}s_{i,1}^{+}s_{i,2}^{+}-\sum_{j\in B}s_{j,1}^{+}s_{j,2}^{+},  \notag \\
Q^{z} &=&\frac{1}{2}\sum_{j\in A\cup B}s_{j,1}^{z}+s_{j,2}^{z},  \label{Q}
\end{eqnarray}%
which are pseudospin operators, obeying the SU(2) Lie algebra, i.e., $%
[Q^{+},Q^{-}]=2Q^{z}$, and $[Q^{z},Q^{\pm }]=\pm Q^{\pm }$. Direct
derivation shows that%
\begin{equation}
\lbrack H,Q^{\pm }]\left\vert \psi _{\pm 0}\right\rangle =0,
\end{equation}%
and%
\begin{equation}
\lbrack \lbrack H,Q^{\pm }],Q^{\pm }]=0,
\end{equation}%
where the ferromagnetic states $\left\vert \psi _{+0}\right\rangle
=\left\vert \Downarrow \right\rangle $ and$\ \left\vert \psi
_{-0}\right\rangle =\left\vert \Uparrow \right\rangle $ are eigenstates of $H
$. Then we obtain a set of degenerate eigenstates of $H$, that is 
\begin{equation}
H(Q^{\pm })^{n}\left\vert \psi _{\pm 0}\right\rangle =0,
\end{equation}%
based on the RSGA. The degeneracy can still be lifted by applying a uniform
external field $B$, $H\rightarrow H+Bs^{z}$, resulting in 
\begin{equation}
H(Q^{\pm })^{n}\left\vert \psi _{\pm 0}\right\rangle =\pm 2Bn\left\vert
\psi _{\pm 0}\right\rangle .
\end{equation}%
These equally-spaced towers are not protected by symmetry and can exhibit
periodic revival phenomena during the dynamical evolution of special initial
states, indicating that they are QMBS. Similar behavior has also
been observed in the Heisenberg model on a Lieb ladder \cite{LiuSun2026}.

Unlike many-body localized phases, this QMBS can occur in translationally
invariant systems without quenched disorder and correspond to atypical
eigenstates embedded within an otherwise thermal spectrum. These states
should typically exhibit anomalously low entanglement entropy, violate the
ETH, and give rise to long-lived coherent dynamics following quenches from
specially prepared initial states \cite{Moudgalya2020Hubbard}.

\section{Entanglement entropy and magnon-pair correlations}

\label{Entanglement entropy and magnon-pair correlations}

It is well-known that entanglement entropy of scar eigenstates grows much
more slowly than in thermal states. In this section, we investigate the
scaling of the entanglement entropy of the eigenstates $\left\vert \psi
_{n}\right\rangle $, given by%
\begin{equation}
\left\vert \psi _{n}\right\rangle =\frac{1}{n!\sqrt{C_{N}^{n}}}%
(Q^{+})^{n}\left\vert \Downarrow \right\rangle ,
\end{equation}%
to determine whether they are quantum scars. We consider a $d$-dimensional\
bipartite lattice with $N=L^{d}$ dimers. In the case with $d=1$, it is
identical to that in Fig. \ref{fig1}(c). We decompose all the $N$\ dimers
into two subsystems $u$\ and $v$, each containing $N/2$\ dimers. The
boundary length $\partial u$\ is~$L^{d-1}$, and the corresponding pair
operators are $Q_{u}^{+}$ and $Q_{v}^{+}$, respectively, satisfying $%
Q^{+}=Q_{u}^{+}+Q_{v}^{+}$.

We perform the Schmidt decomposition of the eigenstates%
\begin{eqnarray}
\left\vert \psi _{n}\right\rangle &=&\frac{1}{\sqrt{\Omega _{n}}}%
(Q_{u}^{+}+Q_{v}^{+})^{n}\left\vert \Downarrow \right\rangle  \notag \\
&=&\sum_{m=0}^{n}\alpha _{m}\left\vert u_{m}\right\rangle \left\vert
v_{m}\right\rangle ,
\end{eqnarray}%
where%
\begin{eqnarray}
\left\vert u_{m}\right\rangle &=&\frac{1}{m!\sqrt{C_{N/2}^{m}}}%
(Q_{u}^{+})^{m}\left\vert \Downarrow \right\rangle _{u},  \notag \\
\left\vert v_{m}\right\rangle &=&\frac{1}{(n-m)!\sqrt{C_{N/2}^{n-m}}}%
(Q_{v}^{+})^{n-m}\left\vert \Downarrow \right\rangle _{v},
\end{eqnarray}%
are the corresponding eigenstates for subsystems $u$ and $v$, with $m\in %
\left[ 0,n\right] $. The factor $\alpha
_{m}=C_{N/2}^{m}C_{N/2}^{n-m}/C_{N}^{n}$~is the Schmidt coefficient. Thus,
the entanglement entropy is

\begin{equation}
S=-\sum_{m=0}^{n}\alpha _{m}^{2}\ln (\alpha _{m}^{2}),  \label{exact}
\end{equation}%
which follows the hypergeometric distribution. In our case, the
hypergeometric distribution can be approximated by the normal distribution
in the thermodynamic limit, then we have%
\begin{equation}
S\approx \frac{1}{2}\ln (2\pi e\frac{n\left( 2N-n\right) }{8N}).
\label{Approximate}
\end{equation}%
A straightforward calculation shows that%
\begin{equation}
S=\frac{d}{2}\ln (L)+\text{const}\propto \ln (\partial u).
\end{equation}%
This reveals that in the thermodynamic limit, the entanglement entropy grows
logarithmically with the boundary length and obeys the subvolume law,
suggesting that the eigenstates may be the quantum scars. Similar results
have been obtained in previous work \cite{Vafek2017, Nakagawa2024}.

Indeed, if we consider a superposition of the degenerate eigenstate set $%
\left\{ \left\vert \psi _{n}\right\rangle \right\} $\ of the following form%
\begin{equation}
\left\vert \phi (\theta )\right\rangle =\sum_{n}d_{n}\left\vert \psi
_{n}\right\rangle ,
\end{equation}%
where $d_{n}=\sqrt{C_{N}^{n}}(-i)^{n}\sin ^{n}(\frac{\theta }{2})\cos ^{N-n}(%
\frac{\theta }{2})$, we have%
\begin{equation}
\left\vert \phi (\theta )\right\rangle =\prod_{j=1}^{N}\left\vert \phi
_{j}(\theta )\right\rangle ,
\end{equation}%
with%
\begin{equation}
\left\vert \phi _{j}(\theta )\right\rangle =\cos (\frac{\theta }{2}%
)\left\vert \downarrow \right\rangle _{j,1}\left\vert \downarrow
\right\rangle _{j,2}-ie^{i\pi j}\sin (\frac{\theta }{2})\left\vert \uparrow
\right\rangle _{j,1}\left\vert \uparrow \right\rangle _{j,2}.
\end{equation}%
It represents a tensor product of the precession states of all local sites,
which indicates that $\left\vert \phi (\theta )\right\rangle $ is a helix
state for nonzero $\theta $. Here, $\theta $ is an arbitrary angle that
determines the profile of the state. When the transformation $H\rightarrow
H+Bs^{z}$\ is applied, $\left\vert \phi (\theta )\right\rangle $ is no
longer an eigenstate of the Hamiltonian. Instead, it evolves as 
\begin{equation}
\left\vert \Phi (t)\right\rangle =\prod_{j=1}^{N}\left\vert \Phi
_{j}(t)\right\rangle ,
\end{equation}%
with%
\begin{equation}
\left\vert \Phi _{j}(t)\right\rangle =\cos (\frac{\theta }{2})\left\vert
\downarrow \right\rangle _{j,1}\left\vert \downarrow \right\rangle
_{j,2}-ie^{i(\pi j-2Bt)}\sin (\frac{\theta }{2})\left\vert \uparrow
\right\rangle _{j,1}\left\vert \uparrow \right\rangle _{j,2},
\end{equation}%
which is also a helix state \cite{Ma2022,Zhang2024}. We note that any two spins in
the state $\left\vert \Phi (t)\right\rangle $\ are correlated. In addition,
even for the state $\left\vert \psi _{n}\right\rangle $, we have%
\begin{equation}
\left\langle \psi _{n}\right\vert
s_{j,1}^{+}s_{j,2}^{+}s_{l,1}^{-}s_{l,2}^{-}\left\vert \psi
_{n}\right\rangle =(-1)^{l-j}\frac{n(N-n)}{N(N-1)},
\end{equation}%
which indicates that $\left\vert \psi _{n}\right\rangle $ also exhibits
ODLRO for finite $n/N$. This shows that state $\left\vert \psi
_{n}\right\rangle $\ describes a condensate state of magnon pairs.

Interestingly, this algebraic construction bears a close resemblance to
Yang's $\eta $-pairing mechanism in the Hubbard model. The $\eta $-pairing
states have recently been recognized as paradigmatic examples of exact scar
eigenstates in interacting fermionic systems. Similarly, the eigenstates $%
\left\{ \left\vert \psi _{n}\right\rangle \right\} $ constructed here
exhibit sub-volume-law entanglement entropy while simultaneously supporting
ODLRO associated with hardcore-boson pair condensation.

\section{Hilbert space fragmentation}

\label{Hilbert space fragmentation}

In this section, we focus on the special case under the resonant condition $%
J=J_{\times }$. We will show that the original Hamiltonian can be mapped
onto a mixed spin-$1$ and spin-$0$ Heisenberg model. This mapping holds
without the restriction of a bipartite lattice. Based on this equivalent
Hamiltonian, we find that the Hilbert space is fragmented. A simple example,
a ladder system, is given as an illustration. Furthermore, numerical
simulations are performed to investigate the system beyond the resonance.

\subsection{Equivalent Heisenberg model}

\label{Equivalent Heisenberg model}

We introduce the local operator%
\begin{equation}
\mathbf{s}_{j}=\mathbf{s}_{j,1}+\mathbf{s}_{j,2},
\end{equation}%
which satisfies the SU(2) commutation relations, i.e., $%
[s_{j}^{+},s_{j}^{-}]=2s_{j}^{z}$, and $[s_{j}^{z},s_{j}^{\pm }]=\pm
s_{j}^{\pm }$. Obviously, the spin value of $s_{j}$ is $0$ or $1$. Based on
identies 
\begin{equation}
s_{j}^{z}=(s_{j,1}^{z}+s_{j,2}^{z}),(s_{j}^{z})^{2}=2s_{j,1}^{z}s_{j,2}^{z}+%
\frac{1}{2},
\end{equation}%
and%
\begin{equation}
s_{j,1}^{+}s_{j,2}^{-}+\text{\textrm{H.c.}}=\mathbf{s}%
_{j}^{2}-(s_{j}^{z})^{2}-1,
\end{equation}%
the original Hamiltonian in Eq. (\ref{H}) under the resonant condition $%
J=J_{\times }$ can be written as%
\begin{eqnarray}
H_{\text{\textrm{eq}}} &=&J\sum_{i\in A,j\in B}s_{i}^{+}s_{j}^{-}+\text{%
\textrm{H.c.}}+J_{\perp }\sum_{j\in A\cup B}\mathbf{s}_{j}^{2}  \notag \\
&&+(\frac{J_{z}}{2}-J_{\perp })\sum_{j\in A\cup B}(s_{j}^{z})^{2},
\end{eqnarray}%
where the constant term is neglected. The equivalent Hamiltonian is a mixed
spin-$1$ and spin-$0$ XY model. A direct derivation shows that 
\begin{equation}
\left[ \mathbf{s}_{j}^{2},H_{\text{\textrm{eq}}}\right] =0,
\label{spin conservation}
\end{equation}%
\begin{figure*}[tbph]
\centering
\includegraphics[width=1.0\textwidth]{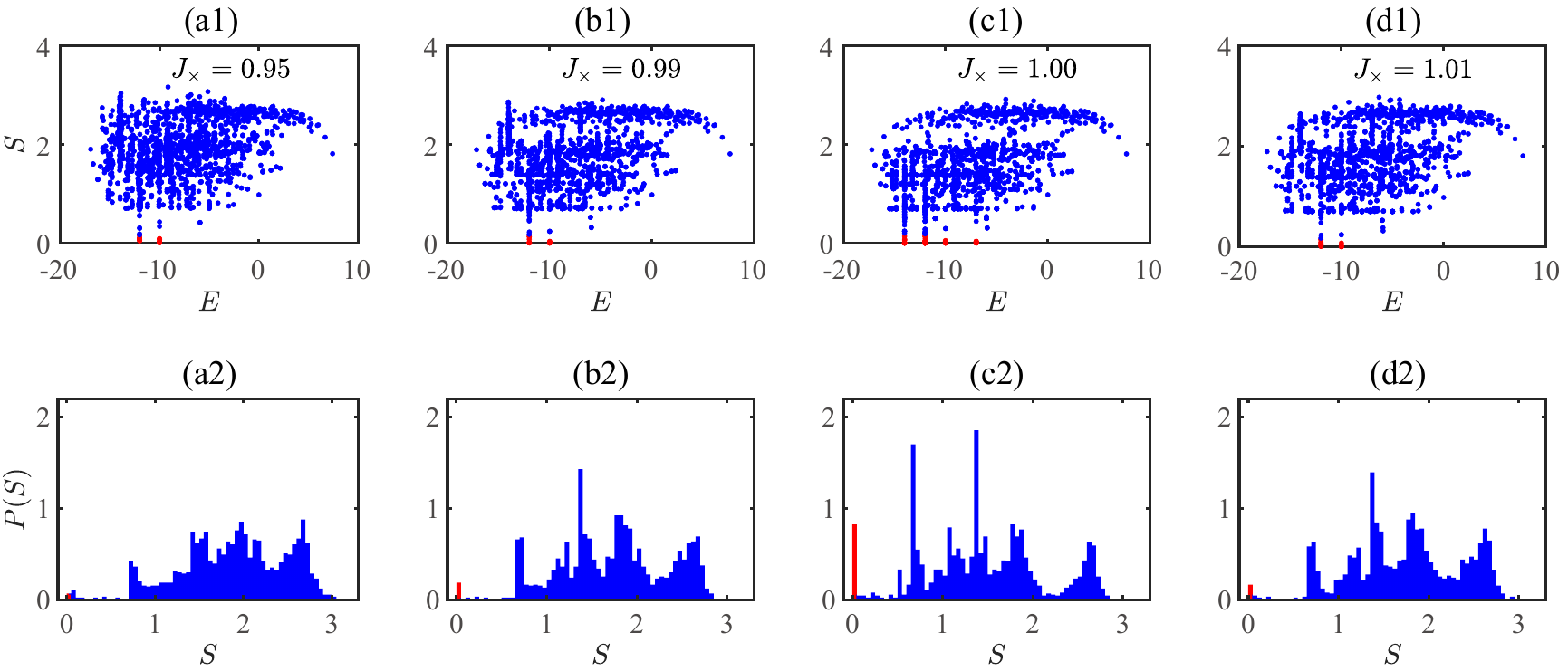} 
\caption{Plots of bipartite entanglement entropy of all eigenstates in the
sector with $\sum_{j=1}^{N}s_{j}^{z}=4$\ and the corresponding probability
distribution for the ladder system [given by Eq. (\ref{H_Ladd})] with $N=8$, $J_{\perp }=5$, and $J_{z}=J=1
$. Panels (a1)--(d1) show the entanglement entropy $S(E)$ as a function of
eigenenergy, while panels (a2)--(d2) display the corresponding probability
distribution $P(S)$. From left to right, the inter-sublattice coupling $%
J_{\times }$ is tuned across the resonance condition $J_{\times }=J$. The
plots with $S\approx 0$\ are indexed by red.\ A pronounced accumulation of
eigenstates with nearly vanishing entanglement entropy emerges at resonance,
signaling strong Hilbert-space fragmentation. Away from resonance, these
low-entanglement states gradually disappear as the fragmented sectors become
connected.}
\label{fig2}
\end{figure*}
which indicates that each possible configuration $\{s_{j}|j\in A\cup B\}$ is
preserved and can be used to index an invariant subspace. We would like to
point out that such a mapping holds even without the restriction to a
bipartite lattice. This may be beneficial for investigating Hilbert space
fragmentation in more complicated systems. As the simplest case, we consider
a system where the layer reduces to a chain. In the invariant subspace with
configuration $\{s_{j}=1|j\in \left[ 1,N\right] \}$, the Hamiltonian $H_{%
\text{\textrm{eq}}}$\ reduces to%
\begin{eqnarray}
H_{\text{\textrm{1D}}} &=&J\sum_{j=1}^{N-1}s_{j}^{+}s_{j+1}^{-}+\text{\textrm{%
H.c.}}+J_{\perp }\sum_{j=1}^{N}\mathbf{s}_{j}^{2}  \notag \\
&&+(\frac{J_{z}}{2}-J_{\perp })\sum_{j=1}^{N}(s_{j}^{z})^{2}.
\end{eqnarray}%
It is a spin-$1$ anisotropic Heisenberg model, which has been studied in
Ref. \cite{Schecter2019}. We note that the result for the constructing exact
eigenstates can be extended to the generalized Hamiltonian in Eq. (\ref{H}).
Notably, a spin-$1$ antiferromagnetic Heisenberg chains exhibit one of the
most celebrated examples of topological quantum matter. In his seminal work,
Haldane predicted that integer-spin chains possess a finite excitation gap
above a unique ground state, in sharp contrast to gapless half-integer-spin
chains \cite{Haldane1983,Haldane1983PRL}. 

\subsection{Hilbert space fragmentation}

\label{HSF}

We note that the coupling between two spins in the Hamiltonian $H_{\text{%
\textrm{eq}}}$ arises from the term $s_{i}^{+}s_{j}^{-}+$\textrm{H.c.},
which vanishes if either of the two spins has spin $0$. The spin state at
such a site is pinned. In this sense, a spin-$0$ site is isolated from the
system. When a set of such pinned spins forms the boundary of a region, the
kinetic constraint for the magnons results in Hilbert-space fragmentation.\
As a simplest case, we consider a system where the layer reduces to a chain.
We find that any spin-$0$\ site can break the chain, providing a kinetic
constraint. Consequently, the Hilbert space fragments into a number of
independent subspaces, and it can be shown that the number of these
subspaces is greater than $2^{N}$, which scales exponentially with the
system size.\ In fact, consider an invariant subspace with a configuration
containing $n$ spin-$0$ sites and $N-n$ spin-$1$ sites. Such a configuration
gives rise to $C_{N}^{n}$ invariant subspaces within this subspace. There
are at least $2^{N}$ invariant subspaces in total,\ resulting in
Hilbert-space fragmentation. Unlike ordinary symmetry sectors, this
fragmentation creates an exponentially many small, dynamically isolated
subspaces. Furthermore, as a special case, when the $N-n$ spin-$1$ sites are
set in fully ferromagnetic states, there are at least $2^{N}$\ tensor
product eigenstates, possessing zero entanglement entropy.\ Take $N=3$ as an
example: the four typical\ eigenstates are $\left\vert 0,0,0\right\rangle $, 
$\left\vert 0,\uparrow ,0\right\rangle $, $\left\vert 0,\uparrow ,\uparrow
\right\rangle $, and $\left\vert \uparrow ,\uparrow ,\uparrow \right\rangle $%
.

\subsection{Entanglement spectrum}

\label{Entanglement spectrum}

When the resonance condition is not satisfied, local spin conservation is
broken, i.e., $\left[ \mathbf{s}_{j}^{2},H\right] \neq 0$. The kinetic
restriction is lifted, and the fragmented subspaces become connected.
Consequently, the tensor product eigenstates may disappear. In the
following, we demonstrate our results in several typical finite-size systems
by numerical simulation. It is well known that quantum entanglement serves
as a probe of thermalization and its breakdown for a given eigenstate.
Presumably, the number of eigenstates with vanishing entanglement entropy
increases significantly when the system approaches the resonance. The
complete set of eigenstates in a fixed $\sum_{j=1}^{N}s_{j}^{z}$\ sector for
a given system is obtained by the numerical diagonalization of the
Hamiltonian. The bipartite entanglement entropy $S(\varepsilon )$\ for the
eigenstate with energy $\varepsilon $ is then computed. The effect of the
Hilbert space fragmentation can be investigated by examining the probability
distribution of $S$, $P(S)$.

In order to reveal the effect of the resonance condition on the Hilbert
space fragmentation, we plot the functions $S(E)$ and $P(S)$ for finite
lattices in Fig. \ref{fig2}. We consider a ladder system with $N=8$,
described by the Hamiltonian given in Eq. (\ref{H_Ladd}). The computation is
performed in the subspace $\sum_{j=1}^{N}s_{j}^{z}=4$. At resonance, there
are $C_{8}^{4}=70$ eigenstates with zero entropy, according to the above
analysis. Other parameters are $J_{\perp }=5$, and $J_{z}=J=1$. Fig. \ref%
{fig2} shows that the entanglement entropy $S$ for a portion of eigenstates
peaks at zero, signaling strong Hilbert-space fragmentation, as expected
when the system is at resonance. In contrast, eigenstates with vanishing
entanglement entropy disappear when the system moves off the resonance.
Before concluding this section, we note that the exact eigenstates
corresponding to magnon-pair condensation persist both at and away from the
resonance point.

\section{Summary}

\label{summary} In summary, we have investigated an anisotropic Heisenberg
model defined on two identical coupled bipartite sublattices. By exploiting
the RSGA, we constructed a family of exact magnon-pair eigenstates and
demonstrated that they form equally spaced towers in the presence of a
uniform magnetic field. These states exhibit off-diagonal long-range order
and sub-volume-law entanglement entropy, indicating their nature as quantum
many-body scar states. We further showed that, under the resonance
condition, the original spin-$1/2$ model can be mapped exactly onto a mixed
spin-$1$ and spin-$0$ XY model. In this representation, the local spin
magnitude becomes a conserved quantity, and spin-$0$ degrees of freedom act
as kinetic constraints that disconnect different dynamical sectors. As a
consequence, the Hilbert space fragments into exponentially many invariant
subspaces. The fragmented structure supports a large number of
low-entanglement eigenstates, including tensor-product states with vanishing
entanglement entropy. Away from the resonance condition, the local
conservation laws are broken and the fragmented sectors become coupled.
Numerical calculations of the entanglement spectrum show that the
accumulation of low-entanglement eigenstates gradually disappears,
consistent with the breakdown of HSF. These results establish a direct
connection between exact scar states and fragmentation phenomena in a broad
class of coupled-spin systems. These findings reveal that the same
underlying spin model simultaneously supports algebraically generated scar
states, long-range correlations associated with magnon-pair condensation,
and kinetically constrained Hilbert space fragmentation. The approach can be
generalized to higher-dimensional lattices and more complicated coupled-spin
systems, offering a promising route toward the systematic construction of
nonthermal eigenstates in interacting quantum many-body models.

\section*{Acknowledgment}

This work was supported by the National Natural Science Foundation of China
(under Grant No. 12374461).

\section*{DATA AVAILABILITY}

The data that support the findings of this article are not publicly
available. The data are available from the authors upon reasonable request.

\bibliography{reference}

\end{document}